\documentclass[aps,prx,twocolumn,nofootinbib]{revtex4-1}
\usepackage{amsmath}
\usepackage{amsfonts}
\usepackage{graphicx}
\usepackage{color}
\usepackage{dsfont}
\usepackage{verbatim}
\usepackage{mathtools}
\usepackage[normalem]{ulem}
\usepackage{comment}
\usepackage{stackrel}
\usepackage{bm}
\usepackage[pdftex]{hyperref}
\hypersetup{colorlinks = true, urlcolor = black, linkcolor = black, citecolor = black}

\definecolor{myblue}{rgb}{.93, .93, 1}

\newcommand{\bsub}{\begin{subequations}}
	\newcommand{\esub}{\end{subequations}}

\begin{document}
	
	\title{Charge density wave and finite-temperature transport in minimally twisted bilayer graphene}

\author{Yang-Zhi~Chou}\email{yzchou@umd.edu}
\affiliation{Condensed Matter Theory Center and Joint Quantum Institute, Department of Physics, University of Maryland, College Park, Maryland 20742, USA}

\author{Fengcheng~Wu}
\affiliation{Condensed Matter Theory Center and Joint Quantum Institute, Department of Physics, University of Maryland, College Park, Maryland 20742, USA}

\author{Jay D. Sau}
\affiliation{Condensed Matter Theory Center and Joint Quantum Institute, Department of Physics, University of Maryland, College Park, Maryland 20742, USA}
		
\date{\today}

\begin{abstract}
We study phenomena driven by electron-electron interactions in the minimally twisted bilayer graphene (mTBLG) with a perpendicular electric field. The low-energy degrees of freedom in mTBLG are governed by a network of one-dimensional domain-wall states, described by two channels of one-dimensional linearly dispersing spin-$1/2$ fermions. We show that the interaction can realize a spin-gapped inter-channel charge density wave (CDW) state at low temperatures, forming a ``Coulomb drag'' between the channels and leaving only one charge conducting mode. 
For sufficiently high temperatures, power-law-in-temperature resistivity emerges from the charge umklapp scatterings within a domain wall. 
Remarkably, the presence of the CDW states can strengthen the charge umklapp scattering and induce a resistivity minimum at an intermediate temperature corresponding to the CDW correlation energy. 
We further discuss the conditions that resistivity of the network is dominated by the domain walls.
In particular, the power-law-in-temperature resistivity results can apply to other systems that manifest topological domain-wall structures.  
\end{abstract}

\maketitle

\section{Introduction}

Twisted bilayer graphene is a paradigmatic example of the moir\'e systems that demonstrate high tunability in the electronic bands. The Dirac Fermi velocity of twisted bilayer graphene can be tuned to zero at certain twist angles (magic angles) \cite{Bistritzer}. Concomitantly, the correlated insulators and superconductors are discovered at low temperatures \cite{tbg1,tbg2}. Subsequent experiments have shown various exotic phenomena including linear-in-temperature resistivity \cite{Polshyn2019,Cao2020PRL} and orbital magnetism \cite{Sharpe2019,Lu2019,Serlin2020}. 
Thanks to these exciting discoveries, moir\'e graphene systems have become exciting platforms for studying strongly correlated phenomena \cite{Codecido2019,Xie2019_spectroscopic,Choi2019,Jiang2019,Chen2019signatures,Burg2019,Kerelsky2019ABCA,Shen2020correlated,Cao2020tunable,Liu2020tunable,Shi2020tunable,Chen2020electrically,Park2020tunable,Hao2020electric}. 

\begin{figure}[t!]
	\includegraphics[width=0.4\textwidth]{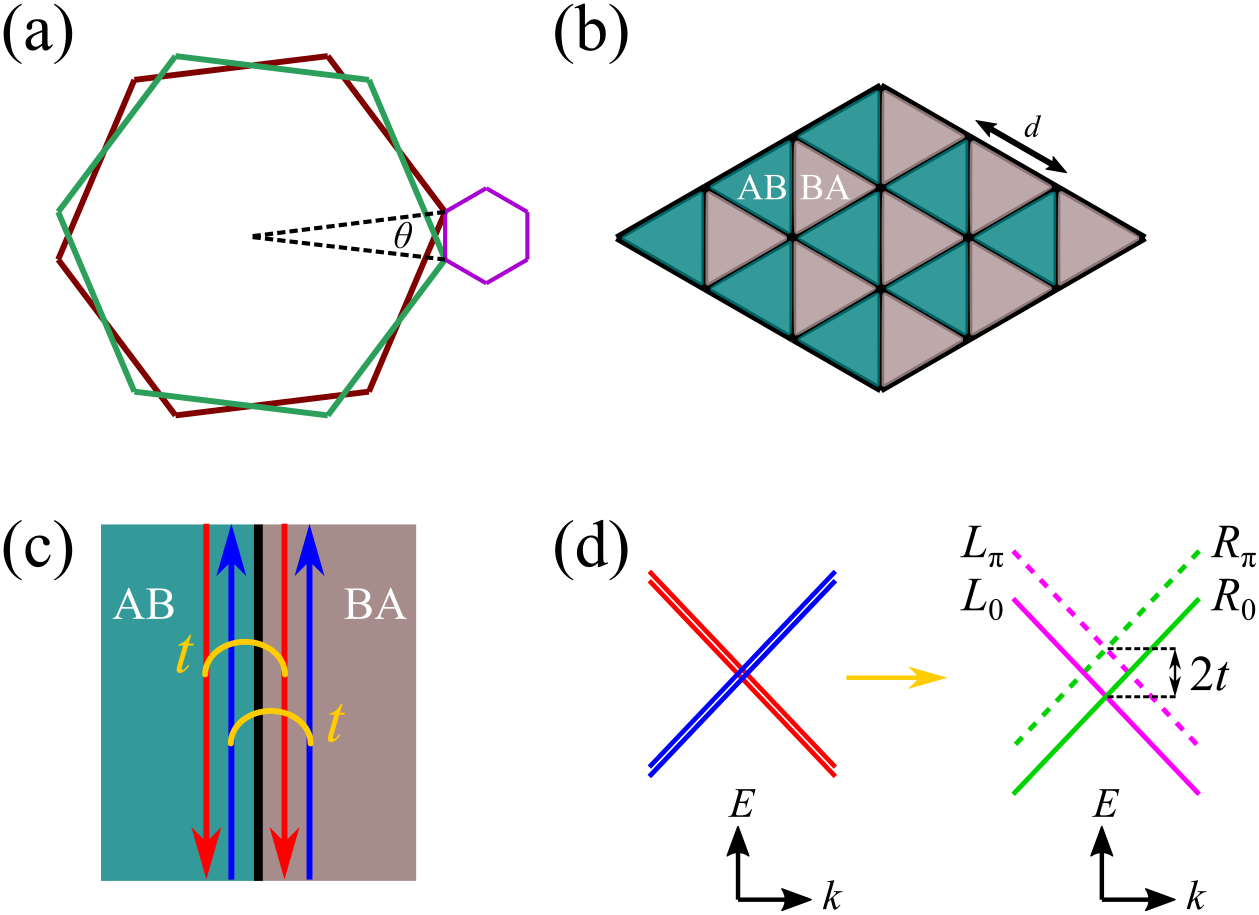}
	\caption{Domain-wall states in mTBLG. (a) The microscopic Brillouin zone (large hexagons) and the moir\'e Brillouin zone (the small hexagon). The size of the moir\'e Brillouin zone depends on the twist angle $\theta$. (b) The overview of mTBLG. $d=a/[2\sin(\theta/2)]$ is the length of one domain-wall segment, where $a$ is the lattice constant of graphene.  
	(c) The electronic degrees of freedom of a domain wall. Blue (red) arrows indicate the microscopic valley K (K$'$) chiral movers. $t$ encodes the hybridization between two chiral movers at the same valley. The spins are not shown for simplicity.
	(d) The degenerate electronic structures are split due to the hybridization. After hybridization, two channels are realized with a relative energy offset $2t$. See main text for detailed discussions.
	}
	\label{Fig:DW_band}
\end{figure}

In addition to the flatband electronic structure at magic angle, the minimally twisted bilayer graphene (mTBLG) with a tiny twist angle ($\ll 1^{\circ}$) can realize a network of one-dimensional (1D) conducting states in the presence of a perpendicular electric field \cite{SanJose2013,Efimkin2018,Yoo2019_reconstruction,Rickhaus2018,Xu2019_GiantOscillation,Huang2018Helical_TBLGexp,Ramires2018,Fleischmann2019,Walet2019,Hou2020,Tsim2020,Chou2020,DeBeule2020,Konig2020,Sunku2020,DeBeule2020_Floquet,Verbakel2021} [see Fig.~\ref{Fig:DW_band}(b)].
The electric field gaps out the low-energy electronic states in AB and BA stacking regions;
the domain walls separating these two regions host gapless 1D conductors, carrying valley-dependent chiralities \cite{Martin2008,Killi2010,Jung2011,Zhang2013,Vaezi2013}. In addition, the AA stacking regions realize junctions that connect domain walls along different directions. The low-energy noninteracting electronic structure in this system can be captured by a phenomenological triangular network model \cite{Efimkin2018}. Novel phases of matters based on the network models are also predicted theoretically \cite{Chou2020,DeBeule2020,DeBeule2020_Floquet}.

Most of the theoretical studies of mTBLG are based on the noninteracting properties of the network model. Meanwhile, the interactions cannot be neglected in 1D systems because the low-energy theory is described by bosonic collective excitations, i.e., the Luttinger liquid theory \cite{Giamarchi_Book}. 
The existing literature on the interacting network models \cite{Wu2019,Chou2019_SC_Int,Chen2020,Konig2020} primarily focuses on the phenomenology in magic-angle twisted bilayer graphene. The role of interaction in mTBLG has not been studied systematically. In particular, The recent transport experiment shows that a low-temperature linear-in-temperature resistivity can emerge in a undoped mTBLG \cite{Xu2019_GiantOscillation}. Can electron-electron interactions realize such an interesting phenomenology?

In this paper, we investigate the triangular domain-wall network model with repulsive electron-electron interactions in the 1D domain walls.
We assume that the valley symmetry is \textit{weakly} broken.
At zero temperature and even in the absence of the valley symmetry, we show that the domain-wall states can develop a spin-gapped inter-channel charge density wave (CDW) state due to the Coulomb interaction among the two channels. Such a CDW state realizes an interlocked fluid among the channels, reminiscent of the 1D Coulomb drag \cite{Klesse2000}.
In addition, the domain-wall state demonstrate a power-law finite-temperature resistivity at sufficiently high temperatures, as a consequence of the charge umklapp scattering in the domain-wall states. The charge umklapp backscattering used in our model takes advantage of the weak breaking of the valley symmetry.
The existence of the inter-channel CDW induces a nonmonotonic temperature dependence in the resistivity, and the temperature of the local resistivity minimum corresponds to the correlation energy of the CDW state. We also discuss the conditions that the the transport of the network is dominated by the 1D domain walls.

The rest of the paper is organized as follows: In Sec.~\ref{Sec:Model}, we introduce a model for the 1D domain-wall states and perform bosonization. The zero-temperature phases are studied in Sec.~\ref{Sec:T=0} where a spin-gapped inter-channel CDW phase is predicted. Then, we calculate the finite-temperature  resistivity with and without the CDW state in Sec.~\ref{Sec:FiniteT}. 
We also discuss the conditions that the network resistivity is dominated by the scattering in the domain walls.
In Sec.~\ref{Sec:Discussion}, the stability of the predictions and other experimental signatures are discussed. In Appendix~\ref{App:Model}, we explain the kinematics in the single-particle Hamiltonian. Detailed discussions on the CDW order parameter and commensurate-incommensurate transition can be found in Appendices~\ref{App:CDW} and \ref{App:SM} respectively. A discussion on the scaling dimension and the high temperature conductivity is given in Appendix~\ref{App:Scaling}. 
We also provide a derivation of the boson self energy in Appendix~\ref{App:SE}, which is related to the finite-temperature resistivity.

\section{Model}\label{Sec:Model}

When two stacked graphene sheets are twisted by a relative angle $\theta$, a moir\'e superlattice emerges with a moir\'e lattice constant $d=a/[2\sin(\theta/2)]$, where $a$ is the lattice constant of the graphene. For $\theta\ll 1^{\circ}$, the lattice relaxation becomes significant, and the AB and BA stacking regions are largely expanded. As a consequence, the mTBLG naturally realizes a periodic arrays of triangular domains as sketched in Fig.~\ref{Fig:DW_band}(b).

Under a sufficiently large out-of-plane electric field, the AB and BA stacking regions in mTBLG become gapped quantum valley Hall (QVH) insulators, and
the domain-wall states separating AB and BA regions are ascribed to the valley Hall kink states \cite{Martin2008,Killi2010,Jung2011,Zhang2013,Vaezi2013}. 
In Fig.~\ref{Fig:DW_band}(b), we consider that the AB (BA) regions realize $\nu=1$ ($\nu=-1$) QVH domains for definiteness. 
As illustrated in Fig. ~\ref{Fig:DW_band}(c), the ``right'' (``left'') mover is inherited from the microscopic valley K (K$'$).
	In the limit $\theta\rightarrow 0$, the momenta of microscopic valleys K and K$'$ are projected to the $k_{\text{1D}}=0$ in each domain-wall link. Thus, the low-energy band of mTBLG is captured by the 1D massless Dirac dispersion. In addition, the difference of the winding number is 2, so each domain-wall state hosts two chiral edge states per spin per valley.

Within the domain-wall states, the single-particle backscattering can be ignored due to the matrix element suppression -- effective valley symmetry \cite{SanJose2013,Efimkin2018}. Such a suppression can be understood by the large momentum transfers between microscopic valleys which makes the overlap of the 2D wavefunctions negligible. Then, the overlap between the right and left mover wavefunctions is parametrically small in a domain wall. Theoretically, the presence of the elastic backscattering can induce gaps in the band structure of the network model.
However, the transport experiments of mTBLG demonstrate no signature of a gap within the network model regime\cite{Rickhaus2018,Xu2019_GiantOscillation}, suggesting that the gap originated from the single-particle backscattering is not experimentally relevant. For simplicity, we ignore the single-particle backscattering completely.

On the other hand, two co-moving states can hybridize without breaking the valley symmetry. 
As a result, the single-particle electronic bands are split by an energy $2t$ where $t$ is the hybridization energy as shown in Fig.~\ref{Fig:DW_band}(d). A detailed discussion is given in Appendix~\ref{App:Model}. The low-energy single-particle Hamiltonian is thus given by
\begin{align}
\label{Eq:H_0_new}\hat{H}_{0}=&v_F\sum_{p,\sigma}\int\! dx\! \left[R^{\dagger}_{p\sigma}\!\left(-i\partial_xR_{p\sigma}\right)-L^{\dagger}_{p\sigma}\!\left(-i\partial_xL_{p\sigma}\right)\right],
\end{align}
where $p=0,\pi$ is the channel index, $\sigma$ denotes the spin, and $R_{p\sigma}$ ($L_{p\sigma}$) is the right (left) mover fermion field.
This noninteracting Hamiltonian contains two channels of 1D spin-$1/2$ linearly dispersing fermions with an energy shift $2t$ as plotted in Fig.~\ref{Fig:DW_band}(c). Correspondingly, the Fermi wavevector difference is $k_{F,0}-k_{F,\pi}=2t/v_F$, independent of the $k_{F,p}$.

In addition to the 1D domain walls, the junctions at the AA stacking regions also play important roles to the network. To simplify our calculation, we consider a network in the decoupled 1D chain limit ($P_{a\bar{a}}=1$ in Ref.~\cite{Chou2020}), corresponding to three parallel arrays of decoupled 1D systems. Then, we consider the effect of junction (i.e., single-particle tunneling between different 1D chains) as a perturbation. Our results also apply to more general situations as long as $\mathcal{C}_{2z}\mathcal{T}$ is preserved. We will discuss the conditions for general network configuration if needed.

In the rest of the section, we discuss the interaction terms and then introduce the bosonization. In particular, for the analysis of transport, we consider backscattering interactions that violate the valley symmetry whereas the single-particle backscattering is ignored completely. 
Note that single-particle backscattering is actually not detrimental to our predictions, but the assumption here simplifies the problem. A detailed discussion is provided in Sec.~\ref{Sec:Discussion}.
Although the bare interacting coupling constants might be small, these interactions can still arise from the renormalization. We will show that the low-energy phases are determined by these interaction backscattering terms, and these interactions can generate the non-Fermi-liquid resistivity phenomenology.

\subsection{Interaction}

We consider short-range repulsive interactions with spin $SU(2)$ symmetry, which are generated either by the screened Coulomb interaction or by renormalization of the low-energy states.
These interactions can be decomposed into the (forward scattering) Luttinger liquid interactions and backscattering interactions (including those violating the valley symmetry). The former can be incorporated by bosonization \cite{Giamarchi_Book,Shankar_Book} in terms of velocities and Luttinger parameters. In particular, we consider an inter-channel density-density interaction given by $V\int dx (n_{0,\uparrow}+n_{0,\downarrow})(n_{\pi,\uparrow}+n_{\pi,\downarrow})$,
where $V>0$ is the strength of interaction and $n_{p\sigma}=R^{\dagger}_{p\sigma}R_{p\sigma}+L^{\dagger}_{p\sigma}L_{p\sigma}$.

Besides the Luttinger interactions, we also focus on the most relevant backscattering interactions that can induce instabilities. 
At first glance, one might argue that the backscattering interactions should be insignificant because of the valley symmetry. It is possible that the bare interaction strengths are parametrically small, but these interactions still arise from renormalization. As we will show in this work, the backscattering interactions can dominate the low-energy phases and can realize a non-Fermi-liquid-like finite-temperature resistivity.
This situation is similar to the helical edge states of the 2D time-reversal topological insulators \cite{Kane2005_1,Kane2005_2,Bernevig2006} -- single-particle backscattering is prohibited while the low-energy phases are controlled by the two-particle interaction backscattering \cite{Wu2006,Xu2006,Kainaris2014,Chou2015,Chou2018}.
Based on the conservation of spin and channel quantum number, we consider the leading backscattering interactions given by
$\hat{H}_{I}=\hat{H}_{I,c}+\hat{H}_{I,s}+\hat{H}_{I,+}+\hat{H}_{I,-}$, where
\begin{subequations}\label{Eq:H_I}
\begin{align}
\label{Eq:H_I_C}\hat{H}_{I,c}
\!=&U_c\sum_{p=0,\pi}\int dx\left[e^{i\delta Q_px}L^{\dagger}_{p\uparrow}R_{p\uparrow}L^{\dagger}_{p\downarrow}R_{p\downarrow}\!+\!\text{H.c.}\right],\\
\label{Eq:H_I_S}	\hat{H}_{I,s}\!=&U_s\sum_{p=0,\pi}\int dx\left[L^{\dagger}_{p\uparrow}R_{p\uparrow}R^{\dagger}_{p\downarrow}L_{p\downarrow}\!+\!\text{H.c.}\right],\\
\label{Eq:H_I_+}\hat{H}_{I,+}
\!=&V_+\sum_{\sigma,\sigma'}\int dx\left[e^{i\delta Q_+x}L^{\dagger}_{0,\sigma}R_{0\sigma}L^{\dagger}_{\pi\sigma'}R_{\pi\sigma'}\!+\!\text{H.c.}\right],\\
\label{Eq:H_I_-}\hat{H}_{I,-}
\!=&V_-\sum_{\sigma,\sigma'}\int dx\left[e^{i\delta Q_-x}L^{\dagger}_{0\sigma}R_{0\sigma}R^{\dagger}_{\pi\sigma'}L_{\pi\sigma'}\!+\!\text{H.c.}\right].
\end{align}
\end{subequations}
In the above expressions, $\delta Q_p=4k_{F,p}-Q$, $\delta Q_{\pm}=2k_{F,0}\pm2k_{F,\pi}-Q$, and $Q=0,\pm 2\pi/d$ is the commensurate wavevector ($d$ is the moir\'e period). In our model, $\delta Q_-=4t/v_F-Q$, independent of $k_{F,p}$.
The interaction $\hat{H}_{I,c}$ ($\hat{H}_{I,s}$) corresponds to the formation of a charge gap (spin gap) in each channel; $\hat{H}_{I,+}$ and $\hat{H}_{I,-}$ are the inter-channel interactions and can lock two channels altogether. For repulsive interactions, all the interaction strengths are positive, i.e., $U_c,U_s,V_+,V_->0$. 
Note that the $U_c$ and $V_+$ terms [Eqs. (\ref{Eq:H_I_C}) and (\ref{Eq:H_I_+}) respectively] do not conserve the valley quantum number, whereas $U_s$ and $V_-$ terms [Eqs. (\ref{Eq:H_I_S}) and (\ref{Eq:H_I_-}) respectively] are valley symmetric.
In general, the interactions $U_c$, $U_s$, $V_+$, and $V_-$ might depend on the $k_F$. We assume them to be $k_F$-independent for simplicity, and our results do not rely on this assumption.

\subsection{Bosonization}

To incorporate the Luttinger liquid interactions, we adopt the standard bosonization \cite{Shankar_Book,Giamarchi_Book}. The right and left mover fields are bosonized to
\begin{align}
R_{p\sigma}=\frac{\hat{\kappa}_{p\sigma}}{\sqrt{2\pi\alpha}}e^{i\left(\phi_{p\sigma}+\theta_{p\sigma}\right)},\,
L_{p\sigma}=\frac{\hat{\kappa}_{p\sigma}}{\sqrt{2\pi\alpha}}e^{i\left(\phi_{p\sigma}-\theta_{p\sigma}\right)},
\end{align}
where $\phi_{p\sigma}$ ($\theta_{p\sigma}$) is the phase-like (phonon-like) boson, $\hat{\kappa}_{p\sigma}$ is the Klein factor, and $\alpha$ is an ultraviolet length scale. The Klein factors are introduced for bookkeeping reasons and can be ignored since they do not affect any of the results in this work.
The long-wavelength density and current operators can be expressed by $n_{p\sigma}=\frac{1}{\pi}\partial_x\theta_{p\sigma}$ and $I_{p\sigma}=-\frac{1}{\pi}\partial_t\theta_{p\sigma}$ respectively.

To study the backscattering interactions, we define the charge and spin bosonic fields in the channel $p$ as follows \cite{Giamarchi_Book}:
\begin{align}
\Phi_{p}^{c}=\frac{1}{\sqrt{2}}\left(\phi_{p\uparrow}+\phi_{p\downarrow}\right),\,\Theta_{p}^{c}=\frac{1}{\sqrt{2}}\left(\theta_{p\uparrow}+\theta_{p\downarrow}\right),\\
\Phi_{p}^{s}=\frac{1}{\sqrt{2}}\left(\phi_{p\uparrow}-\phi_{p\downarrow}\right),\,\Theta_{p}^{s}=\frac{1}{\sqrt{2}}\left(\theta_{p\uparrow}-\theta_{p\downarrow}\right),
\end{align}
where $\Phi_{p}^{c}$ and $\Theta_{p}^{c}$ ($\Phi_{p}^{s}$ and $\Theta_{p}^{s}$) represent the charge (spin) bosonic fields. Incorporating the Luttinger liquid interactions, the Hamiltonian $\hat{H}_0+\hat{H}$ [given by Eqs.~(\ref{Eq:H_0_new}) and (\ref{Eq:H_I})] is bosonized to $\hat{H}_{c}+\hat{H}_{s}+\hat{H}_{cs}$, where
\begin{align}
\nonumber\hat{H}_{c}=&\sum_{p}\int dx \frac{v_c}{2\pi}\left[K_c\left(\partial_x\Phi_{p}^{c}\right)^2+\frac{1}{K_c}\left(\partial_x\Theta_{p}^{c}\right)^2\right]\\
\nonumber&+\frac{2V}{\pi^2}\int dx \left(\partial_x\Theta_{0}^c\right)\left(\partial_x\Theta_{\pi}^c\right)\\
\label{Eq:H_C}&-\frac{U_c}{2\pi^2\alpha^2}\sum_{p=0,\pi}\int dx\cos\left[2\sqrt{2}\Theta_{p}^c+\delta Q_px\right],\\
\nonumber\hat{H}_{s}=&\sum_{p}\int dx \frac{v_s}{2\pi}\left[K_s\left(\partial_x\Phi_{p}^s\right)^2+\frac{1}{K_s}\left(\partial_x\Theta_{p}^s\right)^2\right]\\
\label{Eq:H_S}&+\frac{U_s}{2\pi^2\alpha^2}\sum_{p=0,\pi}\int dx\cos\left[2\sqrt{2}\Theta_{p}^s\right],\\
\nonumber\hat{H}_{cs}=&\frac{-V_+}{\pi^2\alpha^2}\int dx \cos\left[\sqrt{2}\left(\Theta_{0}^c+\Theta_{\pi}^c\right)+\delta Q_+x\right]\\
\nonumber&\,\,\,\,\,\,\times\cos\left(\sqrt{2}\Theta_{s,0}\right)\cos\left(\sqrt{2}\Theta_{s,\pi}\right)\\
\nonumber&+\frac{V_-}{\pi^2\alpha^2}\int dx \cos\left[\sqrt{2}\left(\Theta_{0}^c-\Theta_{\pi}^c\right)+\delta Q_-x\right]\\
\label{Eq:H_CS}&\,\,\,\,\,\,\times\cos\left(\sqrt{2}\Theta_{0}^s\right)\cos\left(\sqrt{2}\Theta_{\pi}^s\right).
\end{align}
In the above expressions\footnote{The minus signs in front of $U_c$ and $V_+$ are due to the bosonization convention used in this work.}, $v_c$ ($v_s$) is the velocity of the charge (spin) bosonic mode, $K_c$ ($K_s$) denotes the Luttinger parameter for the charge (spin) sector, and $V$ encodes the inter-channel Luttinger liquid interaction\footnote{This term is the bosonized form of $V\int dx (n_{0\uparrow}+n_{0\downarrow})(n_{\pi\uparrow}+n_{\pi\downarrow})$.}. In our case with repulsive interactions, $K_c<1$ and $V>0$ are assumed.
$\hat{H}_{c}$ ($\hat{H}_{s}$) is the Hamiltonian for the charge (spin) collective mode and $\hat{H}_{cs}$ describes the inter-channel spin-charge coupling. (Note that spin and charge are decoupled in the absence of $\hat{H}_{cs}$.) Because of the spin $SU(2)$ symmetry, the parameters $v_s$, $K_s$, and $U_s$ are constrained. Under the renormalization group flows with spin $SU(2)$ symmetry, $K_s \rightarrow1$ and $U_s\rightarrow 0$.

For zero-temperature properties, $U_c$ and $V_+$ can be ignored because of lack of momentum conservation at the generic fillings, while $V_-$ can realize an inter-channel CDW state. At finite temperatures, $U_c$ and $V_+$, which break the valley symmetry, can contribute to the resistivity via umklapp mechanism. The rich phenomena driven by interactions will be discussed extensively in the next two sections.

\section{Zero-temperature phases in a domain wall}\label{Sec:T=0}

We study the zero-temperature phase diagram of the 1D interacting domain wall. The Hamiltonian $\hat{H}_c+\hat{H}_s+\hat{H}_{cs}$ [given by Eqs.~(\ref{Eq:H_C}), (\ref{Eq:H_S}), and (\ref{Eq:H_CS})] can be further simplified based on the symmetries and kinematics.

We first examine the spin sector. Since each domain wall possesses spin $SU(2)$ symmetry, the Hamiltonian $\hat{H}_{s}$ [given by Eq.~(\ref{Eq:H_S})] becomes to
\begin{align}\label{Eq:H_S_new}
\hat{H}_{s}'=&\sum_{p}\int dx \frac{v_s}{2\pi}\left[\left(\partial_x\Phi_{p}^s\right)^2+\left(\partial_x\Theta_{p}^s\right)^2\right],
\end{align}
describing a Luttinger liquid of spin with $K_s=1$. 

The charge sector is more complicated. The umklapp interaction $U_c$ in $\hat{H}_{c}$ [given by Eq.~(\ref{Eq:H_C})] can be ignored at zero temperature because $\delta Q_{p=0}\neq 0$ and $\delta Q_{p=\pi}\neq 0$ at generic fillings. 
Meanwhile, $U_c$ is unlikely to induce a CDW near the charge neutrality point ($k_{F,p}\approx0$) for generic network models. To develop a well-define CDW state, it is required that the both correlation length and the CDW wavelength are smaller than the domain-wall length $d$. $k_{F,p}\approx0$ hence contradicts this condition as the CDW wavelength is proportional to $1/k_{F,p}$.
To treat the $V$ term in $\hat{H}_c$ exactly, we introduce another set of new collective charge variables as follows:
\begin{align}
	\Phi_{+}^c=\frac{1}{\sqrt{2}}\left(\Phi_{0}^c+\Phi_{\pi}^c\right),\Theta_{+}^c=\frac{1}{\sqrt{2}}\left(\Theta_{0}^c+\Theta_{\pi}^c\right),\\
	\Phi_{-}^c=\frac{1}{\sqrt{2}}\left(\Phi_{0}^c-\Phi_{\pi}^c\right),\Theta_{-}^c=\frac{1}{\sqrt{2}}\left(\Theta_{0}^c-\Theta_{\pi}^c\right),
\end{align}
where the subscript $+$ ($-$) indicates the symmetric (antisymmetric) collective modes. 
With these new collective variables, the charge sector is described by
\begin{align}
	\nonumber\hat{H}_c'=&\int dx\frac{v_+}{2\pi}\left[K_+\left(\partial_x\Phi_{+}^c\right)^2+\frac{1}{K_+}\left(\partial_x\Theta_{+}^c\right)^2\right]\\
	\label{Eq:H_C_pm}&+\int dx\frac{v_-}{2\pi}\left[K_-\left(\partial_x\Phi_{-}^c\right)^2+\frac{1}{K_-}\left(\partial_x\Theta_{-}^c\right)^2\right],
\end{align}
where $v_+$ ($v_-$) and $K_+$ ($K_-$) are the velocity and the Luttinger parameter of the symmetric (antisymmetric) charge sector respectively. In our case with repulsive interaction, $K_+<K_-<1$ holds generally \cite{Klesse2000}.

Finally, we discuss the inter-channel spin-charge coupling given by $\hat{H}_{cs}$ [Eq.~(\ref{Eq:H_CS})].
The umklapp interaction $V_+$ term can be ignored for generic fillings, similar to the $U_c$ term [Eq.~(\ref{Eq:H_C})].
Therefore, the $V_-$ term is the only important backscattering interaction at zero temperature because the wavevector $\delta Q_-=4t/v_F-Q$ is filling independent and is expected to be small.
The inter-channel spin-charge coupling is reduced to
\begin{align}
\nonumber\hat{H}_{cs}'=&\frac{V_-}{\pi^2\alpha^2}\int dx \cos\left(2\Theta_{-}^c+\delta Q_-x\right)\\
\label{Eq:H_CS_new}&\,\,\,\,\,\,\times\cos\left(\sqrt{2}\Theta_{0}^s\right)\cos\left(\sqrt{2}\Theta_{\pi}^s\right).
\end{align}

In the rest of this section, we study the Hamiltonian $\hat{H}_c'+\hat{H}_s'+\hat{H}_{cs}'$ given by Eqs.~(\ref{Eq:H_C_pm}), (\ref{Eq:H_S_new}), and (\ref{Eq:H_CS_new}). The properties and the stability of the inter-channel CDW phase are discussed.

\begin{figure}[t!]
	\includegraphics[width=0.45\textwidth]{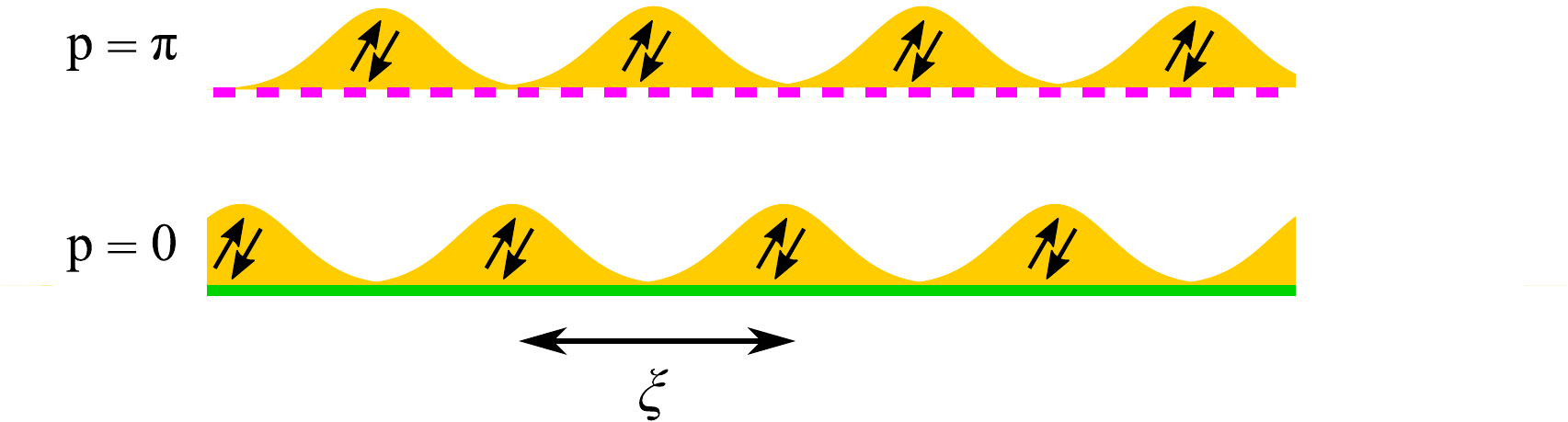}
	\caption{The caricature of the antisymmetric inter-channel CDW state. 
	The green solid line and the magenta dashed line represent channels $p=0$ and $p=\pi$ respectively. The electrons are doubly occupied in each of yellow pocket, and the period of the charge density wave is $\xi=\pi/|k_F|$. The positions of charges are displaced by half of a period among two channels.
	The antisymmetric charge excitation and the spin excitation are gapped.}
	\label{Fig:CDW_a}
\end{figure}

\subsection{Interlocked CDW at $\delta Q_-=0$}\label{Sec:CDW}

For $\delta Q_-=0$ (assuming $k_{F,0}=k_{F,\pi}=k_F$ and $Q=0$), the low-temperature Hamiltonian $\hat{H}_C'+\hat{H}_{S}'+\hat{H}_{CS}'$ is identical to the 1D Coulomb drag problem of two spinful quantum wires \cite{Klesse2000}, and the phase diagram is known.
As long as spin gap in each decoupled channel is absent, the renormalization group flows with the repulsive interactions lead to an infinite $V_-$ quite generally \cite{Klesse2000}. 
In our case, spin gap in each decoupled channel is absent [spin $SU(2)$ symmetry], and $K_-<1$ for repulsive interactions. Therefore, the $V_-$ term is expected to dominate under the renormalization group flows.

The strong coupling fixed point is dictated by Eq.~(\ref{Eq:H_CS_new}) which pins the values of $\Theta_-^c$, $\Theta_{0}^s$, and $\Theta_{\pi}^s$. Thus, the zero-temperature state is described by
a Luttinger liquid in the symmetric charge sector, a gapped state in the antisymmetric charge sectors, and spin gaps for both channels. The proposed CDW state is accompanied by a spin-gap, which has the effect of suppressing the single-particle backscattering that breaks valley symmetry.
Within the bosonization analysis, the leading quasi-long-range order parameter is an antisymmetric inter-channel CDW operator given by
\begin{align}\label{Eq:O_CDW_-}
\hat{O}_-=\sum_{\sigma=\uparrow,\downarrow}\left[\begin{array}{r}
e^{i2k_{F}x}L_{0,\sigma}^{\dagger}R_{0,\sigma}+e^{-i2k_{F}x}R_{0,\sigma}^{\dagger}L_{0,\sigma}\\[2mm]
-e^{i2k_{F}x}L_{\pi,\sigma}^{\dagger}R_{\pi,\sigma}-e^{-i2k_{F}x}R_{\pi,\sigma}^{\dagger}L_{\pi,\sigma}
\end{array}\right]
\end{align}
and $\langle\hat{O}_-(r)\hat{O}_-(0)\rangle\propto\cos\left(2k_F r\right)/r^{K_+/2}$ for $r\gg \alpha$, where $\alpha$ is the ultraviolet length scale. 
The $2k_F$ oscillation suggests that the maximums of the CDW are separated by $\xi=\pi/|k_F|$ (twice of the mean inter-particle distance\footnote{In the electron (hole) doped case, the density of the doped electrons (holes) is $n=2|k_F|/\pi$ corresponding to the mean inter-particle distance $n^{-1}=\pi/(2|k_F|)$.}) and the electrons are doubly occupied as illustrated in Fig.~\ref{Fig:CDW_a}.
A detailed discussion on the inter-channel CDW order parameters can be found in Appendix~\ref{App:CDW}. Intuitively, the zigzag pattern can be understood by the dominating inter-channel Coulomb repulsion, which interlocks two channels altogether. 
A spin gap is developed in this state since it costs a finite energy to create a spin excitation. 
We note that this antisymmetric inter-channel CDW state is similar to the channel singlet state discussed in Refs.~\cite{Wu2019,Konig2020} except that the spin sector is gapped. The discrepancy in the spin sector is due to the different interactions considered in the models.

The magnitude of the CDW correlation can be estimated by the Luttinger liquid analysis \cite{Giamarchi_Book}. Based on the scaling of the $V_-$ term, the correlation energy is 
\begin{align}
\Delta_-\sim\frac{v_F}{\alpha}\left(\frac{V_-}{v_F}\right)^{\frac{1}{1-K_c}},
\end{align}
for $K_c<1$.
The CDW correlation $\Delta_-$ gets smaller when $K_c$ approaches to 1 from below. In mTBLG, the finite size energy scale of a domain wall is given by $v_F/d$ where $d$ is the domain wall length. To have a well-defined CDW state in a domain wall, $\Delta_->v_F/d$ is necessary. 
In addition, the charge period of the interlocked CDW must be much smaller than $d$, hence $d>\xi=\pi/|k_F|$ is required as well. The predicted inter-channel CDW state can exist at generic fillings except for $|k_F|d<\pi$ (close to the charge neutrality point).
We note that the conditions $\Delta_->v_F/d$ and $d>\pi/|k_F|$ are consistent with a spin-gapped CDW state in a finite-size wire of length $d$, and the network formed by such wires is expected to inherit the spin-gapped CDW correlation in each 1D segment.
Although there are two microscopic channels per valley per spin, the inter-channel CDW correlation allows for an effective single channel network description in mTBLG.

\subsection{Commensurate-incommensurate transition for $\delta Q_-\neq0$}

\begin{figure}[t!]
	\includegraphics[width=0.225\textwidth]{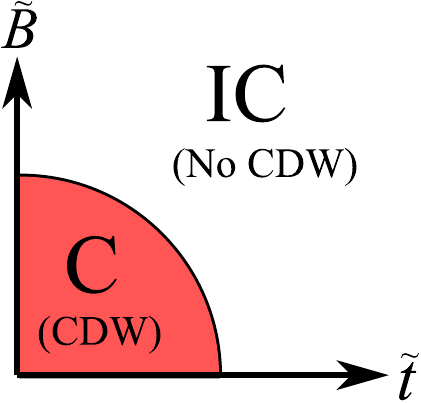}
	\caption{Phase diagram of the commensurate-incommensurate transition. $\tilde{t}=\frac{1}{2}\sqrt{\frac{v_-}{v_FK_-}}\left|\frac{4t}{v_F}-Q\right|\alpha$ and $\tilde{B}=2\mu_BB\alpha/v_F$ are two dimensionless parameters controlling the transition. ($Q$ is the commensurate wavevector, and $\mu_B$ is the Bohr magneton.) The critical line is given by $\tilde{t}^2+\tilde{B}^2=\frac{2V_-}{\pi v_F}$. The red region indicates the commensurate phase (C); the white region represents the incommensurate phase (IC). See Appendix~\ref{App:SM} for a derivation of the phase diagram.
	}
	\label{Fig:C_IC}
\end{figure}

For $\delta Q_-\neq0$, the zero-temperature phase is described by a commensurate-incommensurate transition \cite{PokrovskyTalapov}. The commensurate phase is the same as the $\delta Q_-=0$ limit where the antisymmetric inter-channel CDW is developed. The CDW is absent in the incommensurate phase. Because of the presence of spin gaps in the inter-channel CDW state, a Zeeman field ($B$) can also induce a commensurate-incommensurate transition.
We perform a semiclassical analysis in Appendix~\ref{App:SM}. 
The results are summarized by the phase diagram plotted in Fig.~\ref{Fig:C_IC}.
A sufficiently large $t$ and/or $B$ favor the incommensurate phase and destabilize the inter-channel CDW state, but the phase is essentially the same for generic fillings. We note that the destruction of electron correlation by the Zeeman field is not expected in Ref.~\cite{Wu2019,Konig2020} where spin gapless states are predicted. The predicted Zeeman-field-driven commensurate-incommensurate transition can be examined experimentally.

\section{Finite-temperature transport}\label{Sec:FiniteT}

We study the finite-temperature transport in the network model as realized in mTBLG. 
For network junctions in the decoupled chain limit, the system is described by arrays of 1D systems, and the predictions based on 1D domain walls hold for arbitrary low temperatures.
For generic junction configurations (i.e., away from the decoupled chain limit), there are two distinct temperature regimes in the network model, separated by a crossover temperature scale $T^*=v_F/d$ (where $d$ is the length of the domain wall) \cite{Lee2020}. 
This is equivalent to comparing the thermal wavelength $\lambda_{\text{th}}\sim v_F/T$ and the length of the domain-wall segment.
When the thermal wavelength is much larger than the domain-wall length ($T\ll T^*$), the coherence persists across multiple domain-wall segments, and
the dissipation within a domain wall is negligible. 
In this case, we can view the network model as a 2D Fermi liquid, and
the resistivity in this regime is $\rho(T)\sim A+BT^2$, where $A$ and $B$ are constants. 
On the other hand, when the thermal wavelength is much smaller than the domain-wall length ($T\gg T^*$), ``local equilibriation'' is achieved in each domain wall and the voltage drops are uniformly distributed in the entire system \cite{Lee2020}. In such a situation, the network can be viewed as coupled 1D systems. Thus, the transport of the system is dominated by the umklapp interaction within a domain wall as well as the couplings at the junction. We will focus on the scenario of $T\gg T^*$ in the rest of the section.

In this section, we compute the finite-temperature transport coefficient due to the incommensurate umklapp interactions from valley symmetry breaking in the 1D domain-wall states. To simplify the calculations, we consider the decoupled array limit in the network ($P_{a\bar{a}}=1$ in Ref.~\cite{Chou2020}) and treat the single-particle scatterings at junctions as perturbations. We consider the domain wall with and without the inter-channel CDW correlation.
In particular, a resistivity minimum can develop in the presence of a CDW state. We then discuss the conditions that the network resistivity is dictated by the resistivity of the 1D domain walls.

\subsection{Finite-temperature resistivity in a domain-wall state}

The domain-wall states in mTBLG contains charge, spin, and channel degrees of freedom. Only the channel symmetric charge sector (corresponding to $\Theta_+^c$) contributes to the electric conduction. Strikingly, the presence of inter-channel CDW can still qualitatively change the results as the umklapp interaction is effectively enhanced. 
The primary sources of the finite-temperature resistivity are the
the umklapp interactions $\hat{H}_{I,c}$ and $\hat{H}_{I,+}$. 
For generic fillings, these umklapp interactions are incommensurate (i.e., $\delta Q_p\neq 0$ breaks momentum conservation) and do not impact the zero-temperature phases, and zero resistivity is recovered at zero temperature in a domain wall. 

We consider both the absence and the presence of the inter-channel CDW state which locks two channels. In the former case, we provide an analytical expression of the resistivity due to $\hat{H}_{I,c}$ with the full temperature dependence. In the latter scenario, we perform asymptotic analysis in various limits. The power-law high temperature resistivity and the nonmonotonic behavior due to the CDW are the main predictions in this section.

\subsubsection{Absence of inter-channel CDW}

We first assume that the inter-channel CDW is absent in the domain wall, corresponding to $\Delta_-=0$ or $T>\Delta_-$. In the high temperature limit, the finite-temperature behavior can be obtained by the scaling analysis as discussed in Appendix~\ref{App:Scaling}. The interactions $\hat{H}_{I,c}$ and $\hat{H}_{I,+}$ give rise to $T^{2K_++2K_--3}$ and $T^{2K_+-1}$ resistivity corrections respectively. Those contributions due to $\hat{H}_{I,c}$ and $\hat{H}_{I,+}$ are qualitatively similar except for the precise value in the power-law exponent. To simplify the problem, we consider $\hat{H}_{I,c}$ only and ignore the inter-channel Luttinger liquid interaction $V$ in Eq.~(\ref{Eq:H_C}). ($K_+=K_-=K_c$ and $v_+=v_-=v_c$ in this case.) 

With the simplification mentioned above, the charge sector Hamiltonian is given by $\hat{H}_c=\sum_{p}\hat{h}_{c,p}$,
where $p=0,\pi$ is the channel index and
\begin{align}
\nonumber\hat{h}_{c,p}=&\int dx \frac{v_c}{2\pi}\left[K_c\left(\partial_x\Phi_{p}^{c}\right)^2+\frac{1}{K_c}\left(\partial_x\Theta_{p}^{c}\right)^2\right]\\
\label{Eq:H_cp}&-\frac{U_c}{2\pi^2\alpha^2}\int dx\cos\left[2\sqrt{2}\Theta_{p}^c+\delta Q_px\right].
\end{align}
The finite-temperature transport is limited by the electrons scattering off the incoherent charge fluctuations, described by the $U_c$ term. We note that $\delta Q_p=4k_{F,p}-Q$ with $Q$ being the commensurate wavevector (such as the moir\'e wavevector $2\pi/d$). The momentum relaxation at the moir\'e scale is essential to generate a finite resistivity at finite temperatures and generic fillings. 

Recalling that $\Theta_p^c=\frac{1}{\sqrt{2}}\left(\theta_{p\uparrow}+\theta_{p\downarrow}\right)$, the charge density and current in the channel $p$ are expressed by $n_p=\frac{\sqrt{2}}{\pi}\left(\partial_x\Theta_{p}^c\right)$ and $I_p=-\frac{\sqrt{2}}{\pi}\left(\partial_t\Theta_{p}^c\right)$ respectively. 
Based on the Kubo formula, the dc conductivity (with $e$ and $\hbar$ restored) is expressed by \cite{Chou2015}
\begin{align}
	\label{Eq:dc_cond}\sigma_{\text{1d}}=&-\frac{2}{\pi^2}\frac{e^2}{\hbar}\lim\limits_{\omega\rightarrow 0}\sum_{p=0,\pi}\text{Im}\left[\omega\,\mathcal{G}_p^{(R)}(\omega,k=0)\right],
\end{align}
where $\mathcal{G}_p^{(R)}(\omega,k)$ is the ``dressed'' retarded boson propagator of $\Theta_p^c$. 
The boson self energy is defined through the Dyson equation,
\begin{align}\label{Eq:DysonE}
\left[\mathcal{G}_{p}^{(R)}(\omega,k)\right]^{-1}=\left[G_{p}^{(R)}(\omega,k)\right]^{-1}-\Pi^{(R)}_p(\omega,k),
\end{align}
where $G_{p}^{(R)}$ is the ``noninteracting'' boson propagator, $\Pi^{(R)}_p$ is the self energy correction corresponding to $U_c$, and the superscript $(R)$ denotes to the retarded functions. The decay of the charge mode is related to the imaginary part of $\Pi^{(R)}_p$ and can be characterized by
the inverse scattering length  $\Xi_p$, defined by
\begin{align}\label{Eq:Xi_p_def}
\lim\limits_{\omega\rightarrow 0}\text{Im}\left[\Pi^{(R)}_{p}(\omega,k=0)\right]=-2\omega\Xi_{p}/\pi+O(\omega^2).
\end{align} 
As long as the umklapp interaction $U_c$ term is irrelevant under RG, we expect that $\Xi_p(T\rightarrow 0)\rightarrow 0$. This holds for generic filings and corresponds to a Luttinger liquid phase at $T=0$. Therefore, the leading qualitative features can be captured perturbatively in $U_c$. At the commensurate fillings and $K_c<1$, an interacting gap develops, and a nonperturbative analysis is required to capture the finite-temperature resistivity, which we omit in this work.

With Eqns.~(\ref{Eq:dc_cond}) and (\ref{Eq:Xi_p_def}), the finite-temperature dc resistivity is
\begin{align}
\rho_{\text{1d}}=1/\sigma_{\text{1d}}=&\frac{h}{2e^2}\frac{\Xi_0\Xi_{\pi}}{\Xi_0+\Xi_{\pi}}.
\end{align}
For simplicity, we assume that $\Xi_0\approx \Xi_{\pi}$ and the the resistivity $\rho_{\text{1d}}\propto\Xi_0$. We focus only on generic incommensurate fillings where the umklapp interaction $U_c$ is irrelevant.
The corresponding $\rho_{\text{1d}}$ vanishes at zero temperature which is consistent with the ballistic transport in a Luttinger liquid. 

To extract the inverse scattering length $\Xi_p$, we compute the boson self energy at the second order of $U_c$ analytically. A derivation is sketched in Appendix~\ref{App:SE}. Only the results are presented in the main text.
At the second order in $U_C$, the $\Xi_{p}$ is given by \cite{Sirker2011,Chou2015}:
\begin{widetext}
	\begin{align}\label{Eq:Xi_p}
		\Xi_{p}
		=&
		\left(\tilde{U}_{c} \alpha^{2K_{c}}\right)^2
		\frac{2^{4K_{c}-2}\pi^{4K_{c}}}{v_{c}^{4K_{c}-1}\beta^{4K_{c}-3}}
		\frac{\Gamma\left[1-2K_{c}\right]}{\Gamma[2K_{c}]}\frac{\sin(2 \pi K_{c}) }{\cosh\left(v_{c}\beta \delta Q_{p}/2\right)-\cos(2\pi K_{c})}
		\left|\frac{\Gamma \left[K_{c}+i\frac{v_{c}\beta\delta Q_{p}}{4\pi}\right]}{\Gamma \left[1-K_{c}+i\frac{v_c\beta\delta Q_{p}}{4\pi}\right]}\right|^2,
	\end{align}
\end{widetext}
where $\tilde{U}_{c} = U_{c} / (2 \pi^2 \alpha^2)$ and $\beta=1/T$ is the inverse temperature.
We note that the resistivity derived here from the boson self energy \cite{Oshikawa2002,Sirker2011,Chou2015} is equivalent to the memory function method \cite{Giamarchi1991,Giamarchi_Book}.

\begin{figure}[t!]
	\includegraphics[width=0.35\textwidth]{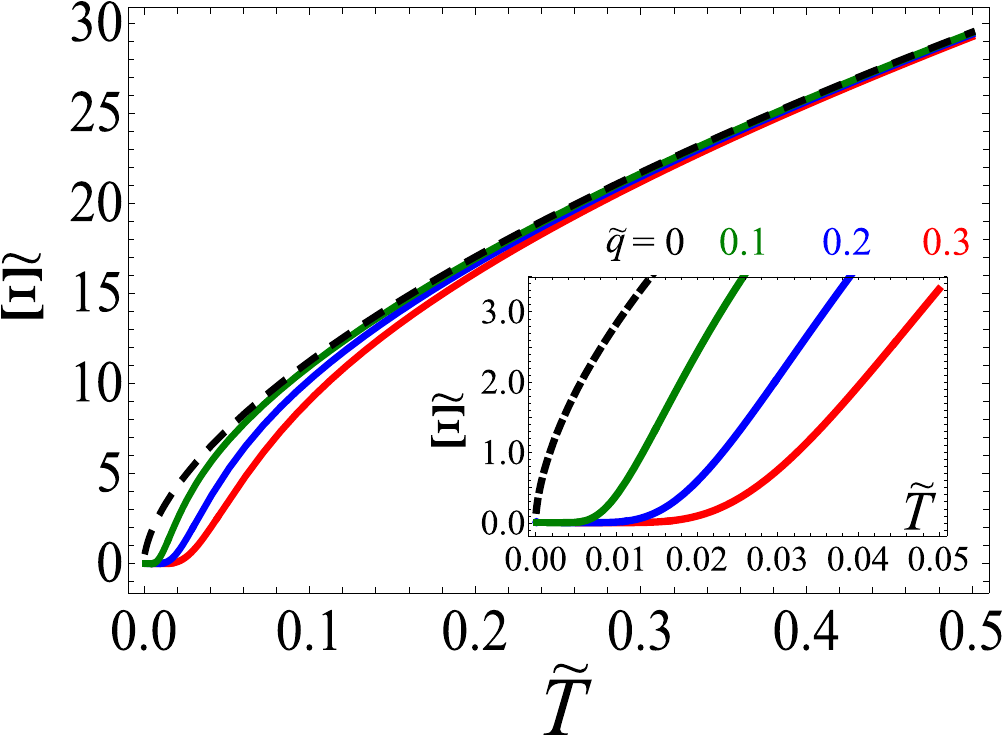}
	\caption{Finite-temperature inverse scattering length in a domain wall (without the inter-channel CDW) based on Eq.~(\ref{Eq:Xi_p}). The resistivity of a domain wall ($\rho_{\text{1d}}$) is proportional to the inverse scattering length.
	For fixed $U_c$ and $v_c$, we plot the dimenionless inverse scattering length, $\tilde\Xi=\Xi_p v_c^2/(\tilde{U}_c^2\alpha^3)$, as a function of the reduced temperature ($\tilde{T}=T\alpha/v_c$) with different values of $\tilde{q}=\delta Q_p\alpha$.
	The black dashed curve represents $\tilde{\Xi}\propto \tilde{T}^{4K_c-3}$, corresponding to $\tilde{q}=0$; the green, blue, and red solid curves correspond to $\tilde{q}=0.1,0.2,0.3$ respectively. $K_c=0.9$ for all the curves.
	For $\tilde{T}>0.3$, all the curves converge and are consistent with $\tilde{\Xi}\propto \tilde{T}^{0.6}$.
	Inset: The low-temperature regime of the inverse scattering length. $\tilde{\Xi}\propto \exp\left(-\frac{\tilde{q}}{2\tilde{T}}\right)$ for $\tilde{T}\ll \tilde{q}$. The onset temperature depends on the value of $\tilde{q}$. The resistivity for $K_c>3/4$ is qualitatively the same as the plot here.
	See main text for a detailed discussion.
	}
	\label{Fig:NFL_R}
\end{figure}

In Fig.~\ref{Fig:NFL_R}, we plot the inverse scattering length $\Xi_p$ as a function of temperature with different values of $\delta Q_p$. 
In the low-temperature limit $(T\ll v_{c} \delta Q_{p})$, the inverse scattering length demonstrates the Arrhenius behavior $\Xi_{p}\propto \exp\left({-\frac{ v_{c}\delta Q_{p}}{2T}}\right)$, recovering the zero resistivity at $T=0$. The low-temperature exponential behavior is due to the phase space restriction ($\delta Q_p\neq 0$) which freezes the umklapp interaction \cite{Giamarchi1991,Giamarchi_Book}. The umklapp scattering is thermally activated at small finite temperatures.
On the other hand, $\Xi_{p}$ gives a power-law behavior in temperature, $\Xi_{p}\propto T^{4K_{c}-3}$, in the limit $T\gg v_{c} \delta Q_{p}$. 
The same power-law exponent can be obtained by scaling analysis as discussed in Appendix~\ref{App:Scaling}.
We note that the qualitative behavior of the finite-temperature resistivity is captured by $\Xi_p$ as $\rho_{\text{1d}}\propto \Xi_p$.
Therefore, the resistivity $\rho_{\text{1d}}(T)\propto \exp\left(-{\frac{ v_{c}\delta Q_{p}}{2T}}\right)$ for $T\ll v_{c} \delta Q_{p}$, and $\rho_{\text{1d}}(T)\propto T^{4K_c-3}$ for $T\gg v_{c} \delta Q_{p}$. We note that the resistivity increases as temperature increases for $K_c>3/4$, while the resistivity decreases as temperature increases for $K_c<3/4$. 

The condition for the power-law finite-temperature resistivity is equivalent to $T\gg v_F k_F$. An important question is if the Luttinger liquid description remains valid for $T\gg v_F k_F$. We answer this in the affirmative. For 1D Dirac bands, the Luttinger liquid theory holds as long as $T>v_F/\alpha\sim W$, where $W$ is the bandwidth. Here, $k_F$ and $W$ are not related to each other. In Fig.~\ref{Fig:NFL_R}, the power-law-in-temperature behavior is clearly shown for $T<v_F/\alpha$.
By contrast, for the 1D quadratic dispersion, the Luttinger liquid description is invalid for $T>v_Fk_F$ as $v_Fk_F$ is also of the same order of the filled bandwidth (i.e., the energy difference between Fermi energy and the bottom of band). Therefore, the power-law finite-temperature conductivity is particular to the systems that realize 1D Dirac dispersion.

\subsubsection{Presence of inter-channel CDW}

In the presence of the inter-channel CDW, only the symmetric charge sector remains gapless. 
In the limit that $T\ll \Delta_-$, the bosonic fields $\Theta_-^c$, $\Theta^s_0$, and $\Theta^s_{\pi}$ are pinned to constant values. Concomitantly, the scaling dimensions of the vertex operators in Eqs.~(\ref{Eq:H_I_C}) and (\ref{Eq:H_I_+}) are reduced, signaling enhancement in these umkalpp scatterings \cite{Ponomarenko2000}.
Based on the scaling analysis (in Appendix~\ref{App:Scaling}), both the $\hat{H}_{I,c}$ and $\hat{H}_{I,+}$ result in $\rho_{\text{1D}}\propto T^{2K_+-3}$ as long as the CDW is well developed. 
The resistivity decreases as temperature increases, qualitatively different from the results without the CDW for $K_c>3/4$. 
This can be understood by a simple mean field decoupling of the umklapp term in the following: In the presence of the inter-channel CDW, the four-fermion terms in Eqs.~(\ref{Eq:H_I_C}) and (\ref{Eq:H_I_+}) can be approximated by
\begin{align}
\nonumber&L^{\dagger}_{p,\uparrow}R_{p,\uparrow}L^{\dagger}_{p,\downarrow}R_{p,\downarrow}\\
\approx&L^{\dagger}_{p,\uparrow}R_{p,\uparrow}\left\langle L^{\dagger}_{p,\downarrow}R_{p,\downarrow}\right\rangle+\left\langle L^{\dagger}_{p,\uparrow}R_{p,\uparrow}\right\rangle L^{\dagger}_{p,\downarrow}R_{p,\downarrow},\\
\nonumber&L^{\dagger}_{0,\sigma}R_{0\sigma}L^{\dagger}_{\pi\sigma'}R_{\pi\sigma'}\\
\approx&L^{\dagger}_{0,\sigma}R_{0\sigma}\left\langle L^{\dagger}_{\pi\sigma'}R_{\pi\sigma'}\right\rangle+\left\langle L^{\dagger}_{0,\sigma}R_{0\sigma}\right\rangle L^{\dagger}_{\pi\sigma'}R_{\pi\sigma'},
\end{align}
where the brackets are replaced by the finite expectation values in the presence of the CDW order [in Eq.~(\ref{Eq:O_CDW_-})]. Thus, the umklapp interactions become more relevant under the mean field decoupling, suggesting an enhancement of the scattering due to CDW.
This mean field approximation also agrees with the scaling analysis based on bosonization in the fermion point (i.e., $K_+=K_-=1$) (see Appendix~\ref{App:Scaling}).
We remark that the 1D CDW is a quasi-long-range order rather than a genuine mean-field like order (i.e., the order parameter fluctuates in a power-law fashion), but the decoupling above provides an intuitive understanding of the interplay between CDW and umklapp interaction.
At sufficiently low temperatures, these umklapp scatterings are frozen because the phase space restriction ($\delta Q_p\neq 0$) creates a Pauli blocking which exponentially suppresses the resistivity.
To fully describe the temperature evolution of resistivity, one needs to take into account the thermal melting of the CDW state. Here, we present results in various asymptotic limit and aim to the qualitative features.
We focus on $\hat{H}_{I,c}$ in the rest of the discussion as the contribution from $\hat{H}_{I,+}$ is qualitative the same. For simplicity, the inter-channel Luttinger interaction is also ignored. Thus, $K_+=K_c$ and $v_+=v_c$.

\begin{figure}[t!]
	\includegraphics[width=0.425\textwidth]{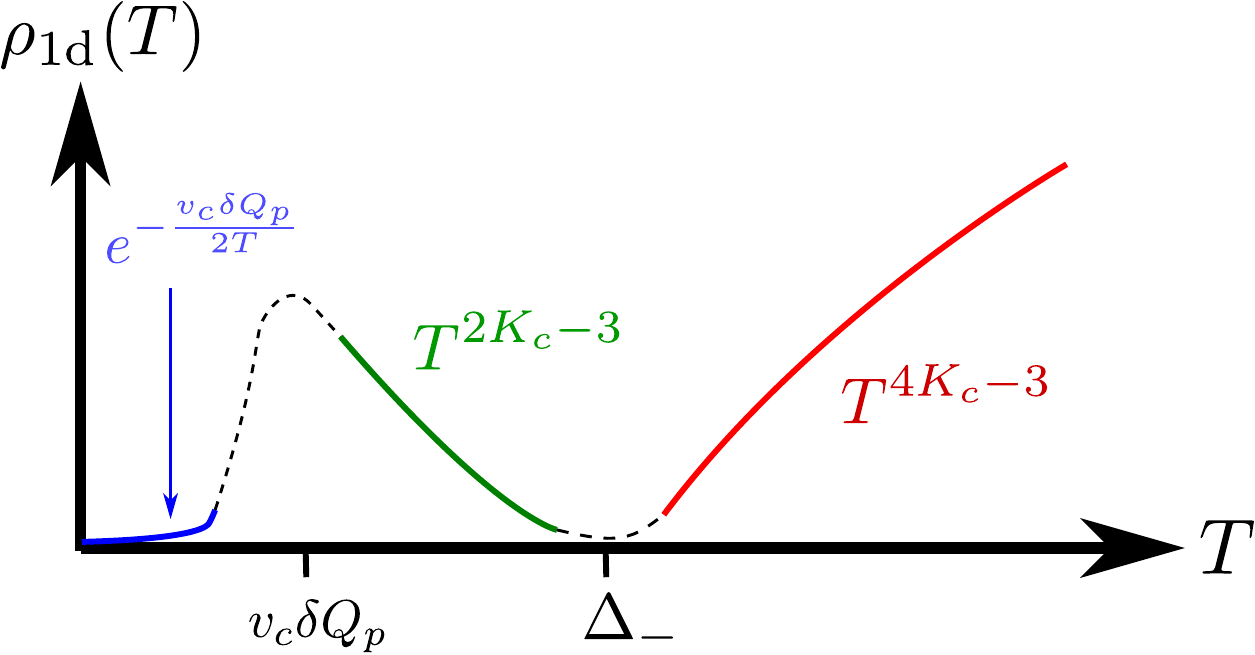}
	\caption{Sketched finite-temperature resistivity in a domain wall state with the inter-channel CDW. In this case, we assume $\Delta_-\gg v_c\delta Q_c$. The blue curve indicates the low-temperature thermal activated behavior for $T\ll v_c\delta Q_p$. The green curve corresponds to a power-law decay $T^{2K_c-3}$ for $v_c\delta Q_p\ll T\ll \Delta_-$. For $T\gg \Delta_-$, the $T^{4K_c-3}$ resistivity is recovered as shown by the red curve. The black dashed lines are used to interpolate different asymptotic regimes. The sketched resistivity is qualitative valid for $3/4<K_c\le 1$.
		See main text for a detailed discussion.
	}
	\label{Fig:RT_CDW}
\end{figure}

Here, we summarize the finite-temperature resistivity in various situations.
The finite-temperature resistivity is qualitatively different from the results without the CDW as long as $\Delta_-\gg v_c\delta Q_p$. 
In this case, a new power-law insulator-like resistivity, $\rho_{\text{1d}}\propto T^{2K_c-3}$, is predicted for $v_+\delta Q_p\ll T\ll \Delta_-$. The insulating temperature dependence here is due to the interplay of the CDW and the umklapp scattering as we discussed previously. 
The low-temperature regime ($T\ll v_c \delta Q_p$) and the high-temperature regime ($T\gg\Delta_-$) give rise to a thermal activated resistivity $\rho_{\text{1d}}\propto \exp\left(-\frac{v_c\delta Q_p}{2T}\right)$ and a power-law resistivity $\rho_{\text{1d}}\propto T^{4K_c-3}$ respectively, similar to the absence of CDW. In Fig.~\ref{Fig:RT_CDW}, we sketch the qualitative behavior of the resistivity versus temperature by interpolating results from the above asymptotic analysis. In particular, for $3/4<K_c\le 1$, a resistivity minimum develops at $T\sim\Delta_-$.
For $K_c<3/4$ and $T\gg v_c\delta Q_p$, the resistivity decreases as temperature increases, and the power-law exponent changes from $2K_c-3$ to $4K_c-3$ at $T\sim\Delta_-$. Thus, it is harder to identify the existence of a CDW state for $K_c<3/4$.
When $\Delta_-< v_c\delta Q_p$, the finite-temperature resistivity is qualitatively the same as the results without the CDW.

\subsection{Non-Fermi-liquid temperature dependence in network}

To recover the network from the 1D domain walls in the decoupled array limit, we include the electron scattering at the junctions. We note that the network model with $\mathcal{C}_{2z}\mathcal{T}$ symmetry is always gapless \cite{Efimkin2018}, and the scattering at the junction cannot induce direct backscattering (a manifestation of the valley symmetry in mTBLG).
For $T\gg T^*=v_F/d$, there are two major sources of resistivity, (i) the inelastic scatterings within a domain wall and (ii) the scatterings at the junctions \cite{Lee2020}. Both (i) and (ii) can contribute to power-law finite-temperature resistivity. However, the contribution from (ii) is temperature independent at $K_c=1$ \cite{Lee2020} while (i) generates a linear-in-temperature behavior at $K_c=1$ for sufficiently high temperatures. 
We focus on the situations where (i) governs the temperature dependence of the resistivity, which is complementary to Ref.~\cite{Lee2020} where the contribution due to (ii) has been studied extensively.

We first consider the limit $\Xi_p(T) d\gg 1$, corresponding to the finite-temperature resistance is much larger than the resistance quanta ($h/e^2$). Since the resistance due to the scattering at the junction is of the order $h/e^2$, the transport of the network model is dominated by the inelastic scattering within a domain wall. This particular limit requires a sufficiently large $U_c$. 
The finite-temperature resistivity shows a non-Fermi-liquid behavior, $\rho(T) \propto T^{4K_c-3}$, at high temperatures, as illustrated in Fig.~\ref{Fig:NFL_R} and \ref{Fig:RT_CDW}. At intermediate temperatures and $K_c>3/4$, the domain walls without the inter-channel CDW show monotonic temperature dependence (Fig.~\ref{Fig:NFL_R}), while the domain walls with the inter-channel CDW demonstrate a nonmonotonic behavior and develop a local minimum around $T\sim\Delta_-$ (Fig.~\ref{Fig:RT_CDW}). $K_c<3/4$ corresponds to a strongly interacting domain wall, which gives rise to an insulator-like resistivity $T^{4K_c-3}$ for sufficiently high temperatures. In such a situation, the existence of the inter-channel CDW is harder to be observed in the finite-temperature resistivity.

For weakly interacting domain-wall states, the Luttinger parameter $K_c\approx 1$. This is likely the situation in the mTBLG experiments as clear signatures of interaction-driven phases have not been reported.
In this limit, the resistance due to scatterings at the junction cannot be ignored. Nevertheless, these contributions are temperature-independent at $K_c=1$. 
As a result, we expect that the major temperature dependence of the resistivity is dictated by the inelastic scattering within a domain wall. At $K_c=1$, we predict a linear-in-temperature resistivity for sufficiently high temperatures. Again, the existence of the CDW can induce a local minimum ($T\sim\Delta_-$) in the resistivity; the resistivity remains to be monotonically increasing in temperature in the absence of the CDW.

In more general cases, the scatterings at the junctions can also contribute to non-Fermi-liquid finite-temperature resistivity (i.e., not $T^2$ resistivity) \cite{Lee2020}. A systematic calculation is needed to fully characterize the finite-temperature resistivity of a network. However, the non-Fermi-liquid resistivity behavior should appear quite generally. The contribution from the $V_+$ term is similar to the the present results due to the $U_c$ term except that the power-law exponent $4K_c-3$ is replaced by $2K_+-1$ in the high temperature regime.

The finite-temperature resistivity discussed here is derived from the umklapp interactions. 
Electron-phonon scattering \cite{Wu2019_phonon} and other mechanisms, which we ignore, also contribute to the finite-temperature resistivity. 
Our predictions are only valid for temperatures smaller than the gap in the AB and BA stacking regions. For temperatures larger than the AB/BA gap, electrons are thermally activated in the AB and BA region, causing an
decrease in the overall resistivity.

\section{Discussion}\label{Sec:Discussion}

We study the phenomena driven by the repulsive electron-electron interactions in the triangular network model as realized in mTBLG with an out-of-plane electric field. We show that the domain-wall states can realize a spin-gapped inter-channel CDW due to the interaction between two microscopic channels. Such a CDW state  can enhance the umklapp interaction and cause a resistivity minimum at a finite temperature.
For sufficiently high temperatures, the umklapp interaction within a domain-wall state can contribute to a non-Fermi-liquid finite-temperature resistivity, $\rho(T)\propto T^{4K_c-3}$. In particular, the linear-in-temperature resistivity is predicted when $K_c=1$.

The results predicted in this work can be examined by the experiments. At zero temperature, we predict a spin-gapped inter-channel CDW state in contrast to the spin-gapless channel singlet states by the previous works \cite{Wu2019,Konig2020}. We expect that the tunneling density of state at Fermi energy is exponentially suppressed in a spin-gapped interlocked CDW state, similar to the prediction for the spin-gapless channel singlet in Ref.~\cite{Konig2020}. This is a manifestation of electron fractionalization (spin, charge, and channel separated) as the domain wall state is actually gapless conducting.
In addition, the 1D spin-gapped inter-channel CDW state can be destabilized by a sufficiently large magnetic field. Therefore, the scanning tunneling microscope with an in-plane magnetic field can test our prediction besides the finite-temperature resistivity discussed in Sec.~\ref{Sec:FiniteT}.
The resistivity generically demonstrate a non-Fermi-liquid behavior, $\rho(T)\propto T^{4K_c-3}$ for sufficiently high temperatures (but the temperatures are still smaller than the gap set by AB/BA domains).
We would like to point out that the observed finite-temperature resistivity in the undoped mTBLG shows a linear-in-temperature behavior below 80K \cite{Xu2019_GiantOscillation}. Our theory reproduces this phenomenology when $K_c=1$. Besides inelastic scattering within the domain wall, the single-particle tunneling at the junction can also realize a non-Fermi-liquid-like resistivity as long as $K_c$ deviates from $1$ significantly \cite{Lee2020}.

Enhancing interaction in mTBLG is preferred to examine our predictions. To this end, one can vary the magnitude of the out-of-plane electric field. Counterintuitively, a smaller electric field results in a stronger interaction \cite{Killi2010}. Meanwhile, a sufficiently strong electric field is necessary to realize the domain-wall network in mTBLG. Therefore, it is optimal to apply an intermediate out-of-plane electric field, such that the velocity is small enough, and the domain-wall states are still sharply defined.

Throughout this work, we have assumed that the single-particle backscattering is absent, but the interaction backscattering can still arise. Such a situation is similar to the interacting helical edges of the 2D time-reversal topological insulator \cite{Wu2006,Xu2006,Kainaris2014,Chou2015,Chou2018}.
In fact, our predictions remain valid even in the presence of the single-particle backscattering as long as the chemical potential is away from the single-particle gaps (due to the single-particle backscattering). In fact, as we have discussed in Sec.~\ref{Sec:CDW}, the presence of a spin-gap in the CDW state suppresses the single-particle backscattering.
The absence of a gap in the experiments \cite{Rickhaus2018,Xu2019_GiantOscillation} might be explained by the disorder smearing effect such that the small single-particle gaps (in the clean limit) disappear. Thus, the absence of single-particle backscattering is a simplification but not a necessary assumption. We also point out that our results are robust against smooth disorder potentials (such as chemical potential fluctuations and the twist-angle disorder \cite{Wilson2020}). The forward scattering disorder (i.e., chemical potential fluctuation) in a domain wall can relax the momentum and modify the finite-temperature resistivity at low temperatures \cite{Fiete2006,Chou2015} while the high-temperature resistivity is qualitatively the same as the clean case. The twist-angle disorder can affect the scatterings at the junction of the domain walls but does not impact a single domain wall significantly. Thus, we expect that our theory can apply to other systems manifesting conducting network systems as well.
In particular, the predicted non-Fermi-liquid finite-temperature resistivity should be relevant to the minimally twisted double bilayer graphene \cite{Kerelsky2019ABCA} and networked topological helical surfaces \cite{Chou2019}.

\section*{Acknowledgments} 

This work is supported by the Laboratory for Physical Sciences (Y.-Z.C. and F.W.), by JQI-NSF-PFC (supported by NSF
grant PHY-1607611, Y.-Z.C. and J.D.S.), and NSF DMR1555135 (CAREER, J.D.S.)

\appendix

\section{Model with two channels}\label{App:Model}

As illustrated in Fig.~\ref{Fig:DW_band}(b), the electrons in a domain-wall are described by 
\begin{align}
	&\nonumber\hat{H}_{0}\\
	\nonumber=&v_F\!\sum_{j,\sigma}\!\int\! dx\! \left[\psi^{\dagger}_{K,j\sigma}\!\left(-i\partial_x\psi_{K,j\sigma}\right)-\psi^{\dagger}_{K',j\sigma}\!\left(-i\partial_x\psi_{K',j\sigma}\right)\right]\\
	\nonumber&-\mu\sum_{j,\sigma}\int\! dx\!\left[\psi^{\dagger}_{K,j\sigma}\psi_{K,j\sigma}+\psi^{\dagger}_{K',j\sigma}\psi_{K',j\sigma}\right]\\
	\label{Eq:H_0_old}&-t\sum_{\sigma}\int\! dx\!\left[\psi^{\dagger}_{K,2\sigma}\psi_{K,1\sigma}+\psi^{\dagger}_{K',2\sigma}\psi_{K',1\sigma}+\text{H.c.}\right]
\end{align} 
where $j=1,2$ is the chiral mover index, $\sigma=\uparrow,\downarrow$ denotes the spin, $v_F$ is the fermi velocity, $\mu$ is the chemical potential, $t$ encodes the hybridization of two edge states, and $\psi_{K,j\sigma}$ ($\psi_{K',j\sigma}$) indicates the field operator for the $j$th edge electron of valley $K$ (valley $K'$) with spin $\sigma$. In Eq.~(\ref{Eq:H_0_old}), the 1D counter-propagating massless Dirac fermions carry well-defined valley quantum number, and the inter-valley single-particle scattering is absent. The hybridization $t$ term describes the symmetry-allowed intra-valley tunneling between two co-moving fermions.

The hybridization $t$ between two edges is crucial to the electronic structure in a domain wall. 
Similar to the studies for the two-leg ladder problem \cite{Giamarchi_Book}, we introduce $R_{0\sigma}$ and $L_{0\sigma}$ ($R_{\pi\sigma}$ and $L_{\pi\sigma}$) representing the bonding (antibonding) fermionic fields. These fermionic fields are
defined as follows: $R_{0\sigma}=\frac{1}{\sqrt{2}}\left(\psi_{K,1\sigma}+\psi_{K,2\sigma}\right)$, $L_{0\sigma}=\frac{1}{\sqrt{2}}\left(\psi_{K',1\sigma}+\psi_{K',2\sigma}\right)$,
$R_{\pi\sigma}=\frac{1}{\sqrt{2}}\left(\psi_{K,1\sigma}-\psi_{K,2\sigma}\right)$, and $L_{\pi\sigma}=\frac{1}{\sqrt{2}}\left(\psi_{K',1\sigma}-\psi_{K',2\sigma}\right)$. To eliminate the chemical potential terms, we perform linear transformations, $R_{p\sigma}(x)\rightarrow e^{ik_{F,p}x} R_{p\sigma}(x)$ and $L_{p\sigma}(x)\rightarrow e^{-ik_{F,p}x} L_{p\sigma}(x)$, where $k_{F,0}=(\mu+t)/v_F$ and $k_{F,\pi}=(\mu-t)/v_F$.
With the above transformations, equation~(\ref{Eq:H_0_old}) becomes to Eq.~(\ref{Eq:H_0_new}).

\section{CDW order parameters}\label{App:CDW}
When $V_-$ term [Eq.~(\ref{Eq:H_CS_new})] becomes dominating, the Luttinger liquid is unstable to the formation of certain quasi-long-range order.
Because of the nature of $V_-$ (equivalent to the Coulomb drag problem), we focus on quasi-long-range ordered states of charge. 
There are two possible inter-channel CDW order parameters, characterized by symmetric and antisymmetric linear combinations of single channel CDW states, given by
\begin{subequations}
\begin{align}
\hat{O}_+=&\sum_{\sigma=\uparrow,\downarrow}\left[\begin{array}{r}
	e^{i2k_{F}x}L_{0,\sigma}^{\dagger}R_{0\sigma}+e^{-i2k_{F}x}R_{0\sigma}^{\dagger}L_{0\sigma}\\[2mm]
	+e^{i2k_{F}x}L_{\pi\sigma}^{\dagger}R_{\pi\sigma}+e^{-i2k_{F}x}R_{\pi\sigma}^{\dagger}L_{\pi\sigma}
\end{array}\right],\\
\hat{O}_-=&\sum_{\sigma=\uparrow,\downarrow}\left[\begin{array}{r}
e^{i2k_{F}x}L_{0,\sigma}^{\dagger}R_{0\sigma}+e^{-i2k_{F}x}R_{0\sigma}^{\dagger}L_{0\sigma}\\[2mm]
-e^{i2k_{F}x}L_{\pi,\sigma}^{\dagger}R_{\pi\sigma}-e^{-i2k_{F}x}R_{\pi\sigma}^{\dagger}L_{\pi\sigma}
\end{array}\right].
\end{align}
\end{subequations}
$\hat{O}_+$ and $\hat{O}_+$ contain only the $2k_F$ components of charge density operators. The inter-channel CDW order parameters are bosonized to
\begin{subequations}
\begin{align}
\hat{O}_+=&\frac{2}{\pi \alpha}\left[\begin{array}{r}
\sin\left(\sqrt{2}\Theta_{0}^{c}+2k_Fx\right)\cos\left(\sqrt{2}\Theta_0^{s}\right)\\[2mm]
+\sin\left(\sqrt{2}\Theta_{\pi}^{c}+2k_Fx\right)\cos\left(\sqrt{2}\Theta_{\pi}^{s}\right)
\end{array}\right],\\
		\hat{O}_-=&\frac{2}{\pi \alpha}\left[\begin{array}{r}
			\sin\left(\sqrt{2}\Theta_{0}^{c}+2k_Fx\right)\cos\left(\sqrt{2}\Theta_0^{s}\right)\\[2mm]
			-\sin\left(\sqrt{2}\Theta_{\pi}^{c}+2k_Fx\right)\cos\left(\sqrt{2}\Theta_{\pi}^{s}\right)
		\end{array}\right].
\end{align}
\end{subequations}

When $V_-\rightarrow \infty$, to represent the ground state, we can choose $\Theta_{-}^c(x)=\pi/2+n\pi$, $\Theta_{0}^s=\sqrt{2}n_0\pi$, and $\Theta_{\pi}^s=\sqrt{2}n_{\pi}\pi$, where $n$, $n_0$, and $n_{\pi}$ are integers. Other solutions that minimize the $V_-$ term are equivalent. The CDW order parameters become 
\begin{align}
\hat{O}_+\rightarrow 0,\,\,\hat{O}_-\rightarrow\frac{4}{\pi\alpha}\cos\left(\Theta_+^c+2k_Fx\right).
\end{align}
The symmetric inter-channel CDW order parameter vanishes exactly when $V_-\rightarrow \infty.$ Moreover, the equal-time correlation function of $\hat{O}_-$ is given by
\begin{align}
\nonumber\left\langle\hat{O}_-(r)\hat{O}_-(0)\right\rangle=&\frac{4}{\pi^2\alpha^2}\left\langle e^{i2k_Fx}e^{i\left(\Theta_+^c(r)-\Theta_+^c(0)\right)}\right\rangle+\text{H.c.}\\
\approx&\frac{8}{\pi^2\alpha^2}\cos\left(2k_Fr\right)\left(\frac{\alpha}{r}\right)^{K_+/2}.
\end{align}
The antisymmetric CDW state is decaying with a power-law exponent $K_+/2$ and oscillates in space. 
The oscillation period $\xi=\pi/k_F$ indicates the charge pocket separation in the CDW state. Since $\hat{O}_-$ is anti-correlated in the channel space, the two CDW states of different channels are offset by half of the oscillation period. These features are illustrated in Fig.~\ref{Fig:CDW_a}.

Besides the CDW states, spin density wave order and superconducting order parameters can be constructed \cite{Giamarchi_Book}. However, these order parameters are either zero or subleading, consistent with the physical intuition for the repulsively interacting two channel problem. 

\section{Semiclassical analysis for $\delta Q_-\neq 0$}\label{App:SM}

To construct a phase diagram with both $\delta Q_-$ and the applied Zeeman field, we perform a semiclassical analysis in this section. 
Neglecting the decoupled $\Theta_+$ sector, the semiclassical energy functional (ignoring $\Phi$'s) is given by
\begin{align}
	\nonumber \mathcal{E}=&\int dx\frac{1}{2\pi}\left[\frac{v_-}{K_-}\left(\partial_x\Theta_{-}^c\right)^2+v_F\sum_{p=0.\pi}\left(\partial_x\Theta_{p}^s\right)^2\right]\\
	\nonumber&+\frac{V_-}{\pi^2\alpha^2}\int dx \cos\left(2\Theta_{-}^c+\delta Q_-x\right)\\
	\label{Eq:CM_energy}&\,\,\,\,\,\,\times\cos\left(\sqrt{2}\Theta_{0}^s+\delta Q_sx\right)\cos\left(\sqrt{2}\Theta_{\pi}^s+\delta Q_sx\right),
\end{align}
where $\delta Q_s={2\mu_B B}/{v_F}$, $\mu_B$ is the Bohr magneton, and $B$ is the Zeeman field. 
Then, we follow the analysis used in Ref.~\cite{Hanna2001}. When the $V_-$ term is negligible, the energy is govern by the derivative terms, favoring constant values of $\Theta_{-}^c$, $\Theta_{0}^s$, and $\Theta_{\pi}^s$. Thus, the free energy $\mathcal{E}=\mathcal{E}_{0}=0$, corresponding to the incommensurate solution. In the opposite limit, the commensurate solution can be constructed by minimizing the cosine term ($V_-$ term). We choose $\Theta_{-}^c(x)=\pi/2-\delta Q_- x/2$, $\Theta_{0}^s=\Theta_{\pi}^s=-\delta Q_s x$. The other solutions that minimize the cosine term are equivalent.
The energy for the commensurate solution is given by
\begin{align}
	\mathcal{E}_{1}=L\left[\frac{v_-}{8\pi K_-}\delta Q_-^2+\frac{v_F}{2\pi}\delta Q_s^2-\frac{V_-}{\pi^2\alpha^2}\right],
\end{align}
where $L$ is the length of the 1D system. 

For $\mathcal{E}_1<\mathcal{E}_0=0$, the ground state is in the commensurate phase indicating the formation of an inter-channel CDW state.
For $\mathcal{E}_1>\mathcal{E}_0$ , the incommensurate phase is energetically favored. 
A phase diagram based on the above analysis is plotted in Fig.~\ref{Fig:C_IC}.
Recalling that $\delta Q_-={4t}/{v_F}-Q$ and $\delta Q_s={2\mu_B B}/{v_F}$. A sufficiently large $t$ and/or $B$ favors the incommensurate phase and destabilizes the inter-channel CDW state. In particular, the destruction of CDW by the Zeeman field can be examined experimentally.

\section{Scaling dimensions in high temperature limit}\label{App:Scaling}

For sufficiently high temperatures, the conductivity in a domain wall can be estimated by a Fermi's golden rule argument. We consider an inelastic interaction as follows:
\begin{align}
	\mathcal{S}_I=g\int d\tau dx\, \mathcal{O}(\tau,x),
\end{align}
where $g$ is the coupling constant and $\mathcal{O}$ is the interaction operator.
At the second order perturbation theory, the conductivity due to an interaction is
\begin{align}
	\sigma_{\text{1d}}\sim\frac{1}{g^2}\frac{1}{T^r},
\end{align}
where $T$ is the temperature and $r$ is the temperature exponent. 

We determine the exponent $r$ by a dimensional analysis.
The conductivity in one dimension has the scaling dimension $[\sigma_{\text{1d}}]=-1$, the temperature has the scaling dimension $[T]=1$, and $[g]=-[\mathcal{O}]+2$.
With these scaling dimensions, we conclude that 
\begin{align}
r=-2[g]+1.
\end{align}
The temperature exponents obtained in this way are consistent with the results in the high temperature limit of the Kubo conductivity.

\subsection{Without CDW}

In the absence of the inter-channel CDW, the scaling dimensions of the bosonized operators given by the $U_c$ and the $V_+$ interactions are
\begin{align}
&\left[\cos\left(2\sqrt{2}\Theta_{0,\pi}^c\right)\right]=\left[\cos\left(2\Theta_+^c\pm 2\Theta_-^c\right)\right]=K_++K_-,\\
&\left[\cos\left(2\Theta_+^c\right)\cos\left(\sqrt{2}\Theta_0^s\right)\cos\left(\sqrt{2}\Theta_{\pi}^s\right)\right]=K_++1,
\end{align}
where we have omitted the incommensurate wavevectors. This analysis is valid for temperatures much larger than the scale set by the incommensurate wavevector. One can easily check that the $r=2K_++2K_--3$ for the $U_c$ interaction and $r=2K_+-1$ for the $V_+$ interaction.

\subsection{With CDW}

In the presence of the inter-channel CDW, the scaling dimensions of the operators are modified. Assuming the CDW correlation is infinitely strong, one can replace $\Theta_-^c$, $\Theta^s_0$, and $\Theta^s_{p}$ by constant values corresponding to minimizing Eq.~(\ref{Eq:H_CS_new}). The modified scaling dimensions of the bosonized operators given by $U_c$ and $V_+$ interactions are
given by
\begin{align}
	&\left[\cos\left(2\sqrt{2}\Theta_p^c\right)\right]=\left[\cos\left(2\Theta_+^c\right)\right]=K_+,\\
	&\left[\cos\left(2\Theta_+^c\right)\cos\left(\!\sqrt{2}\Theta_0^s\right)\cos\left(\sqrt{2}\Theta_{\pi}^s\right)\right]=	\left[\cos\left(2\Theta_+^c\right)\right]=K_+.
\end{align}

In this case, $r=2K_+-3$ for both the $U_c$ and $V_+$ interactions.

\section{Derivation of Self energy}\label{App:SE}

To derive the retarded boson self energy,
we formulate the problem in the imaginary-time path integral. Since two channels are decoupled in Eq.~(\ref{Eq:H_cp}), we focus on the channel $p$ in this appendix.
The \textit{axial} action (after integrating out $\Phi_{p}^c$) is given by
$\mathcal{S}=\mathcal{S}_{c}+\mathcal{S}_{I}$, where
\begin{subequations}
\begin{align}
	\mathcal{S}_{c}=&\int d\tau dx\frac{1}{2\pi v K_c}\left[\left(\partial_{\tau}\Theta_{p}^c\right)^2+v^2\left(\partial_{x}\Theta_{p}^c\right)^2\right],\\
\mathcal{S}_{I}=&-\frac{U_c}{2\pi^2\alpha^2}\int d\tau dx\cos\left[2\sqrt{2}\Theta_{p}^c+\delta Q_px\right],
\end{align}
\end{subequations}
where $\tau$ denotes the imaginary time. The self energy can be derived via the effective action as discussed in Ref. \cite{Peskin_Book}. The expression of the space-time self energy is 
\begin{align}
\nonumber&\Pi_p(\tau_1,\tau_2;x_1,x_2)\\
\label{Eq:Pi_p_x_t}=&-4\tilde{U}_c^2\!\left\{\!\begin{array}{c}
\!\cos\left(\delta Q_p(x_1-x_2)\right)e^{8\left[G_p(\tau_1-\tau_2,x_1-x_2)-G_p(0,0)\right]}\\[2mm]
\!-\delta_{1,2}\int d\tau_3dx_3\cos\left(\delta Q_px_3\right)e^{8\left[G_p(\tau_3,x_3)-G_p(0,0)\right]}
\end{array}\!\!
\right\},
\end{align}
where $\tilde{U}_c=U_c/(2\pi^2\alpha^2)$, $\delta_{1,2}$ denotes to $\delta(\tau_1-\tau_2)\delta(x_1-x_2)$, and $G_p$ is the noninteracting boson propagator. 

The vertex function $e^{8\left[G_p(\tau,x)-G_p(0,0)\right]}$ is crucial to the self energy and is expressed by
\begin{align}
\nonumber&e^{8\left[G_p(\tau,x)-G_p(0,0)\right]}\\
=&\left\{\frac{\left(\frac{\pi\alpha}{v_c\beta}\right)^2}
{\sinh\left[\frac{\pi}{v_c\beta}\left(x+iv_c\tau\right)\right]\sinh\left[\frac{\pi}{v_c\beta}\left(x-iv_c\tau\right)\right]}
\right\}^{2K_c}
\end{align}
with $\beta=1/T$ being the inverse temperature. Then, we perform Fourier transform and obtain
\begin{widetext}
	\begin{align}
		\mathcal{F}(i\omega_n,k)=&\int d\tau dx\,e^{i\omega_n\tau-ikx}e^{8\left[G_p(\tau,x)-G_p(0,0)\right]}\\
		=&2^{4K_c-2}\left(\frac{\pi\alpha}{\beta v_c}\right)^{4K}\frac{v_c\beta^2}{\pi}\frac{\sin(2K_c\pi)}{\pi}\frac{\Gamma\left[1-2K_c\right]\,\Gamma\left[1-2K_c\right]\Gamma\left[K_c+\frac{1}{2}(\mu+\nu)\right]\Gamma\left[K_c+\frac{1}{2}(\mu-\nu)\right]}{\Gamma\left[1-K_c+\frac{1}{2}(\mu+\nu)\right]\Gamma\left[1-K_c+\frac{1}{2}(\mu-\nu)\right]},
	\end{align}
\end{widetext}
where $\Gamma$ denotes the Gamma function, $\mu=\frac{\beta\omega_n}{2\pi}$, and $\nu=i\frac{\beta v_ck}{2\pi}$.

The self energy [Eq.~(\ref{Eq:Pi_p_x_t})] in the Fourier space can be expressed in terms of $\mathcal{F}$ as follows:
\begin{align}
\tilde{\Pi}(i\omega_n,k)=-\frac{4}{2}\tilde{U}_c^2\left[\begin{array}{c}
\mathcal{F}(i\omega_n,k+\delta Q_p)+\mathcal{F}(i\omega_n,k-\delta Q_p)\\[2mm]
-\mathcal{F}(0,\delta Q_p)-\mathcal{F}(0,-\delta Q_p)
\end{array}
\right].
\end{align}
To obtain the retarded self energy defined in Eq.~(\ref{Eq:DysonE}), we perform analytic continuation $i\omega_n\rightarrow \omega+i0^+$. Finally, one can derive Eq.~(\ref{Eq:Xi_p}) based on the retarded self-energy and Eq.~(\ref{Eq:Xi_p_def}).





\end{document}